\documentclass[pra,nopacs,floatfix,reqno,twocolumn,a4,footinbib]{revtex4}
\usepackage{graphicx}

\begin{document}

\title{Employing real experiments and modern viewpoints in the teaching of modern physics}
\author{Muhammad Sabieh Anwar}
\altaffiliation{Department of Physics, Syed Babar Ali School of Science and Engineering,
Lahore University of Management Sciences (LUMS), Lahore 54792, Pakistan}
\email{sabieh@lums.edu.pk}
\homepage{http://physlab.lums.edu.pk}
\thanks{I like to thank the colleagues and students who helped me in delivering the course on modern physics and building the experiments in the spring semester of 2013: Dr. Ata-ul-Haq, Dr. Imran Younus, Shama Rashid, Muddassir Moosa, Ali Hasan, Muhammad Yousaf, Ahmed Zubairi, Rabiya Salman, Amrozia Shaheen, Hafiz Rizwan, Ali Hasan, Junaid Alam, Hasaan Majeed, Maryan Khaqan, Faran Irshad, Qurrat-ul-Ain and Hasnain Ali.}

\date{\today}
\begin{abstract}
This is a report of a course on modern physics designed and taught to undergraduate science and engineering students in the Spring of 2013. The course, meant for freshmen, attempts to integrate statistical mechanics into non-classical physics and introduces some novel teaching ideas such as the field approach in contrast to typical wave and particle viewpoints traditionally covered in usual textbooks. The various modern applications of quantum theory in realizing practical devices are recounted in a way that is amenable to beginners. Especially, we describe the inclusion of visually appealing and carefully designed robust experiments inside the formal classroom environment. This collection of applications and demonstrations serves as a useful collection of teaching aids and can be easily transformed into active learning exercises. The impact of this teaching strategy on student learning and its role in exciting an interest in physics is assessed.
\end{abstract}

\maketitle

\section{Motivation}

The contents, style and format of the year-long compulsory physics stream for science and engineering students have been subject to several revisions~\cite{transform}. While many universities worldwide and in particular North American universities teach two separate and at times disconnected courses on ``mechanics'' and ``electricity and magnetism'', here in our university (LUMS) we experimented instead with two four-credit hour courses on ``mechanics'' and ``modern physics'', while one fourth of the latter would comprise ideas from thermal physics.

The underlying philosophy was to expose \emph{all} students to trends in contemporary physics. Generally, at the end of the conventional sequence, non-physics majors walk away effectively knowing nothing about lasers, light, nuclear radiation, subatomic particles, spins, cosmology, astrophysics, semiconductors or electronic devices. We therefore decided to swap the order of ``modern physics'' and ``electricity and magnetism'' and also incorporate some thermal physics into modern physics. I attempted to perform this integration seamlessly so that thermodynamics and quantum concepts do not remain segregated with their distinct vocabularies, rather they were welded together in a unified framework. In fact Chabay and Sherwood's work presented an excellent way forward in this regard~\cite{chabay1}. These authors have also produced a textbook~\cite{chabay2} that infuses the atomistic viewpoint right into the heart of mechanics and electromagnetism. In their book, thanks to a carefully deigned sequence and well chosen examples, one finds electrons, atoms, molecules, stars, black holes and many other ideas from modern physics interwoven with Newtonian concepts.

In this article, my aim is to mention the largely non-traditional approach followed in delivering this calculus-based course to a class of 225 students. In Section~\ref{sec:organize}, I outline the organization of the course and how I assessed student performance---the kinds of questions asked and the kind of material provided to students. In Section~\ref{sec:contents}, the central ideas discussed in the course are mentioned, especially the unconventional ones. This is followed by a brief recounting of the applications and in-class demonstrations built for this course. These are the topics of Sections~\ref{sec:applications} and \ref{sec:demonstrations}. I have also gathered some responses from the student feedback and have captured a useful metric highlighting student performance in Section~\ref{sec:feedback}. Finally, I conclude in Section~\ref{sec:conclusions}.

\section{Course organization and assessment tools\label{sec:organize}}

The course was composed of lectures, recitations, tutorials and office hours whose break-up is shown in Table~I. The central topics were covered in the main lectures where the entire class gathered. The practical demonstrations and quizzes were conducted inside the lecture time. Students were divided into four roughly equal sections for the recitations which covered illustrative problems and exercises through an interactive rapport. For tutorials, conducted by advanced undergraduate students, the class was divided into six sections enabling closer interaction with the class. These tutorials were based on specially designed conceptual exercises or delved into the mathematical mechanics of assigned home work problems. For examples, students were taught how to solve a differential equation, or were provided with the necessary tools in complex numbers, plotting of trigonometric functions and so on. These tools could not be adequately addressed in the main lectures. Finally, during office hours, the class freely approached the instructors and student teaching assistants with their learning inquiries.

\begingroup
\squeezetable
\begin{table}[!t]
\begin{tabular}{lll}
\hline
Nature of contact & Duration in minutes & Teacher \\
  \hline
  Main lectures & 2 $\times$ 75 min, 1 $\times$ 50 min & lead instructor \\
  Recitations & 1 $\times$ 75 min & co-instructors \\
  Tutorials & 1 $\times$ 60 min & students and lead instructor\\
  Office hours & 3 $\times$ 60 min & students and lead instructor  \\
  \hline
\end{tabular}
\label{table:course-organization}
\caption{Outline of the course organization.}
\end{table}
\endgroup

Students were assessed on the basis of quizzes, home works, mid and final term exam. Home works were of two kinds: the usual ``individual'' home works that were attempted by each student independently and the ``collaborative'' home works in which students teamed up in groups of four. These home works dealt with technically advanced applications and the concepts were tested in sequentially ordered exercises. For example, in one homework students were expected to understand the concept of thermal equilibrium from the predicting of fluctuations away from equilibrium. The homework was derived from Prentis's article~\cite{prentis} and required students to calculate, for example, ``how long, on average, in a time of one hour, will a person hover above the earth's surface at a height of $1~$cm (assuming different numerical values of Boltzmann's constant)?'' A student commented on these collaborative home works, \begin{quote}``There were collaborative assignment [sic.], which fosters camaraderie and team work among peers and felt more effective than individual homework assignments.''\end{quote} Some sample questions from the various instruments are reproduced in the appendix while a complete set can be obtained from the author's website~\cite{website}.

\section{Overview of the course contents\label{sec:contents}}
\subsection{Integrated statistical thermodynamics}

 The unusual sequence of topics ensured the gelling of thermal concepts with the structure of matter. No doubt statistical mechanics provided the correct framework. The course started with the harmonic oscillator while parallel recitation sessions taught students basic techniques in solving ordinary differential equations. From harmonic oscillators the concept of an Einstein solid was built and the meaning of internal energy was elaborated. Thermal energy was precisely defined as the random undirected component of the internal energy. Then straight away, I delved into the quantization of internal energy, followed by the first law of thermodynamics $\Delta U=Q+W+\text{``other forms''}$ where $Q$ and $W$ were respectively defined as microscopic and macroscopic forms of work and ``other forms'' included chemical, electric energies and so on. With the help of entropy I progressed towards the second law of thermodynamics highlighting the viewpoints of maximization of entropy as well as the equitable distribution (partition) of internal energy. I ended up paying considerable attention to entropy and demystifying the many misconceptions circulating the meaning of this term. I found the idea of energy of energy dispersal very useful. I presented several examples, using the rules of probability, of finding out the ways of distributing energy amongst quantized levels composing an Einstein solid. This approach led to the numerical computation of entropies for large (as well as nano-sized) systems and hence connecting the microscopic quantized descriptions with macroscopic properties. I think that bridging the gap between micro and macro interpretations of entropy was valuable and highlighted the purely physical origins of entropy, thereby bringing this poorly understood quantity out of the realm of mere abstraction. In this respect, Schroeder's beautiful article~\cite{schroeder1} and the text~\cite{schroeder2} proved to be wonderful references as well as the series of Leff's articles so accurately titled \emph{Removing the mystery of entropy and thermodynamics}~\cite{leff1,leff2,leff3,leff4,leff5}.

\subsection{Waves, particles and fields}

Traditional courses introducing quantum physics focus on waves and particles and emphasize the duality. An electron sometimes behaves as a ``wave'' and sometimes as a ``particle'' with the resulting implication that a particle and wave become the ``real'' entities and these quantum objects appear to change their identity depending upon how we choose to look at them. Particles become associated with ``matter'' and ``waves'' with ``radiation'' and the physics teacher then presents examples of when the particles ``act like'' waves (e.g. interference) and when the ``waves'' act like particles (e.g. when the entities interact and share momentum and energy such as upon a photon hitting the eye's retina).

A pedagogically satisfying (and hence different) and an all-embracing conceptual framework was provided by introducing the concept of a quantum field~\cite{hobson0,hobson00}. In this strategy, the Schrodinger's wavefunction $\Psi(\mathbf{r},t)$ is simply stated as the nonrelativistic limit of a quantum field. This field is sometimes called a ``matter wave'' or an ``electromagnetic field'' depending on what we are talking about, electrons, neutrons, bucky balls, viruses, or light. (For an alternative viewpoint criticizing the excessive reification of quantum fields, see~\cite{mermin}).

Let's talk about the electron. In the teaching approach advocated by Hobson, the electron \textit{is} a quantized field. By adopting this viewpoint in my teaching I discovered some useful implications. Represented by $\Psi(\mathbf{r},t)$ the field extends throughout space and hence is non-local. When we draw the field at some time $t_0$, we are actually sketching a portrait of the (amplitude of the) field in space but frozen in time $\Psi(\mathbf{r},t_0)$. The single electrons fills up all of space. This reasoning easily explains how the electron can get across nodes~\cite{nelson} without resorting to twisted artificial arguments like ``tunneling across nodes''.

The field concept has great unifying power. It is equally descriptive of, say, electrons and light. Hence the distinction between matter and radiation fades away and both are treated symmetrically on equal footing. An interference pattern is formed by the superposition of the components of the field without any pressing need for declaring that the electron is ``behaving'' like a wave.

What then does the word ``particle'' mean? In Hobson's viewpoint, the ``quantum'' in the ``quantum field'' is attributable to the word ``particle'' in traditional parlance. The field holds energy, just like the electromagnetic field from a radio transmitter carries energy which is transported to the receiver that plays the music. When the field interacts locally, it deposits, fully or partially, its energy and momentum onto something else. For example, suppose a beta particle, a field, impacts a Geiger Muller tube. This is a local interaction---local because the impact occurs at one point inside extended space. The energy inside the field bunches up and transmits to the detector, producing an audible click or flash of light. This interaction is called a ``particle''!
Finally what's ``quantum'' about all this? Noting that the energy in the audible signal produced from impact will be discrete multiples of some unit, we recognize that the energy in the field is also quantized, $n$ electrons will produce sound that is $n$ times as loud.

\subsection{The uncertainty principle ``turned around"}

The uncertainty principle is the cornerstone of quantum physics and took a lot of my teaching time. While the principle is generally considered to be a lifeless elucidation of the barren inequality $\Delta x \Delta p \geq \hbar/2$ and is skimmed over cursorily in traditionally taught physics courses, eliciting its inner meaning requires a specially nuanced and careful teaching approach~\cite{hobson1}.

Here the role of proper explanatory language is brought to the fore~\cite{williams}.  Ironically, I felt that the word ``uncertainty" itself became an impediment in an understanding of the uncertainty principle. Students (and the laity) often use the negative connotations inherent in ``uncertainty" to lull themselves into the restrictive pessimism that ``there is a limit to what you can know; and after all, \emph{God does indeed play dice}".  This dangerous attitude---which makes quantum physics ad consequently the whole of physics into an inchoate farrago of fuzziness, unpredictability and weakness---can  be turned around by redefining ``uncertainty" as a ``realm of possibilities"~\cite{hobson2}.

What then does a finite $\Delta x$ mean? It signifies that a quantum object can show up within a range of locations when measured by an appropriate location measuring apparatus. An identically prepared object might as well end up at a different location (if $\Delta x \neq 0$) when detected with the same apparatus under totally identical conditions or multiple runs of the identically initialized experiment.

I aided this interpretative sleight with a series of lectures focusing on the archetypal single photon interference experiment and its highly intriguing modern variants. I felt that exposing students straight away to a real experiment allowed them to have better grasp of the frequently used terms ``randomness", ``unpredictability", ``uncertainty" and the ``realm of possibilities". This converted theoretical subtleties and syntactic nuances into concrete physical occurrences. For this purpose, I introduced two kinds of measurement schemes: (a) ``wait and sweep" and (b) ``wait". Here is what I meant by these.

\subsubsection{The ``wait and sweep" experiment (style A)}

The ``wait and sweep" strategy uses one tightly focused telescope which I defined as a ``detector in a state of waiting''. The telescope tightly observes one small region on the detection screen also called the field of view. A photon issuing from the source inside a double-slit arrangement may reach the telescope's field of view which is triggered in the process, or it may land somewhere else and the event goes undetected. Now finding  the complete realm of possibilities, $\Delta x$, necessitates the telescope to focus on one region, wait and then change its orientation, focusing on some other region. Ultimately, the ``wait and sweep" strategy will generate a histogram (with location along the abscissa and the number of location-dependent triggers along the ordinate). The spread of this histogram determines $\Delta x$.

Whether the telescope triggers or not is connected with randomness but the probability of triggering can be determined from first principles. To carry the argument further, the sheer choice between firing and no-firing makes the event ``unpredictable" but since the probability is computable the event is also deterministically ``predictable". This is how \emph{quantum uncertainty} can be defined for students and distinguished from the everyday usage of the word ``uncertainty''. No wonder, a careful choice of words and a tightly woven description went a long way in comforting perplexed students who initially saw all of this as a labyrinth.
\begin{figure}[!here]
\begin{center}
\includegraphics[scale=0.6]{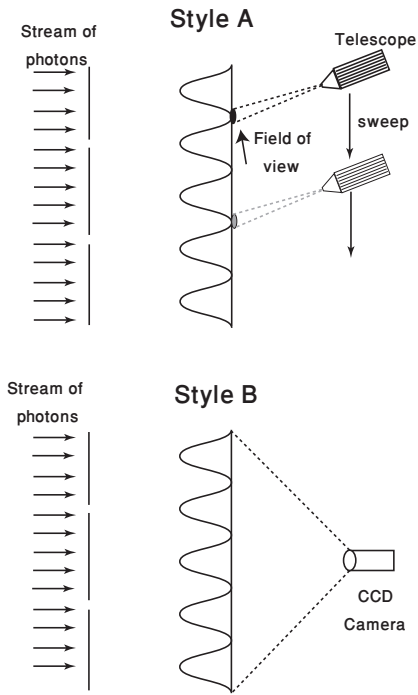}
\caption{The two variants of the double-slit interference experiment described in the text.
\label{quantum1}}
\end{center}
\end{figure}

\subsubsection{The ``wait" experiment (style B)}

The second experimental strategy is to simply ``wait" and employs a wide-angle CCD camera whose field of view is the \emph{entire} detection screen. Individual photons register on different locations and as time ensues, a distribution of markings builds up on the screen. Each marking is an imprint of a single photon hitting the screen. In this scheme, there is absolutely no need to physically reorient the camera. The spatter of photons directly generates a histogram, showing that identically prepared photons end up depositing their energies on different locations and the spread of the electron imprints is measure of $\Delta x$. Since each photon gives one and only one blip, the deposited energy in each blip being the same, we say that the field is quantized. Fortuitously a recent experiment by Bach \textit{et al.} (of style B) achieved the direct implementation of an interference pattern with electrons~\cite{bach}. We could watch the pattern build up in real time, thanks to the authors who had made the video from their detection camera public, something I delightedly showed in class. Enabled by these experimental schemes, here are a few points I was able to highlight.

\begin{enumerate}
\item
A photon hitting the screen at $x_0$ does not imply that the photon really existed at $x_0$, for another identical photon may end up somewhere else. Detection at $x_0$ is one out of a ``realm of possibilities". This realm is represented by a ``probability distribution".

\item
Focusing a telescope on a slit placed before the screen changed the outcome on the screen. We say that the obtainment of ``which-path" information erases interference.

\item
I also motivated a distinction between destructive and non-destructive measurements by comparing experiments with photons and electrons, respectively. Photons triggering telescopes are annihilated in the process while electrons may shed off only portions of their energy without being annihilated. A home work designed around this theme can be downloaded from my website~\cite{website}.
\end{enumerate}

\subsection{Uncertainty diagrams}

I also came to recognize the use of uncertainty diagrams as potent teaching aids. Even though similar diagrams are frequently used in describing quadrature states and squeezing of light~\cite{fox}, I was introduced to their pedagogical value  through Hobson's work~\cite{hobson2}. I believe such diagrams produce remarkably clear yet condensed depictions of realms of possibilities of conjugate variables, $\Delta x$ and $\Delta p$, as well as their expectation values $\bigl<x\bigr>$ and $\bigl<p\bigr>$. The collapse entailed by a strong projective measurement also has a straightforward depiction on an uncertainty diagram.

\begin{figure}[!h]
\begin{center}
\includegraphics[scale=0.45]{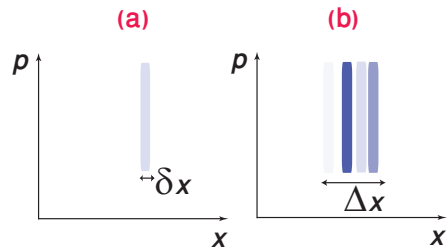}
\caption{Uncertainty diagram generated (a) as a single measurement on the location of an electron, $\delta x$ being the instrumental uncertainty. (b) The diagram builds up in multiple experiments and the quantum uncertainty (called the realm of possibilities) $\Delta x$ shows up.
\label{quantum2}}
\end{center}
\end{figure}

What can we say about the instrumental uncertainty arising out of, for example, the finite resolution of the scale of the measuring instrument or  the resolution of an optical instrument (called the diffraction limit)? These are questions normally overlooked with the downside that silence often leaves room for the creeping in of incorrect self-assurances. The superficial student of quantum physics is content with casually declaring that ``the quantum uncertainties cannot be reduced by better instruments or the skill of the experiment". This statement is indeed valid but it's also important to emphasize that in a \textit{single measurement}, it is well nigh possible that the measurand is measured to an instrumental precision tighter than its realm of possibilities. For example in a single shot, a location measuring scale could assign the object's location to within a region much smaller than $\Delta x$. It is only when the measurement is repeated on identically prepared copies that the probability distribution builds up and the quantum realm of possibilities $\Delta x$ becomes evident. A careful exposition of uncertainty diagrams as they build up in repeated experiments also throws light on the interplay between instrumental and quantum uncertainties quite nicely. One may see Figure~\ref{quantum2} in this regard.

\section{Converting quantum `myths' to `realities': sample applications\label{sec:applications}}

A significant portion of the course was decidedly devoted to modern applications and devices. Choosing what to teach in this section presented considerable difficulty as students evidently had no idea of solid state and atomic physics nor electronics. Besides I wanted to choose comprehensive, relevant but comprehensible examples that exposed that quantum physics has an inevitable place in modern micro- and nanoelectronics. I wasn't satisfied with the tunnel (Leo Esaki's) diode which adorns most textbooks because these are specialized devices hardly very difficult to buy. I ended up with the single electron transistor (SET), field effect transistor (FET), FLASH memory and of course, the scanning tunneling microscope.

\subsection{Greeting electrons, one at a time}

\begin{figure}[!t]
\begin{center}
\includegraphics[scale=0.4]{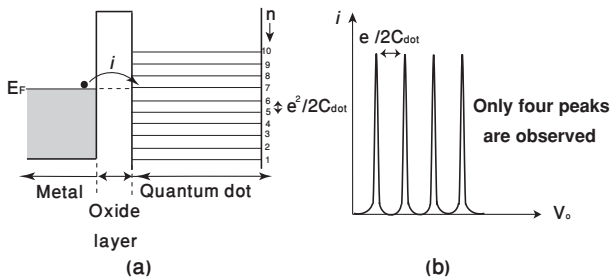}
\caption{(a) Energy level diagram of a single electron transistor and (b) graph showing its quantized conductance.
\label{myth33}}
\end{center}
\end{figure}

Amenable discussions of the SET can be found in Rogers \emph{et al}.'s Nanotechnology~\cite{rogers}  and Kastner's review articles on artificial atoms published in Physics Today~\cite{kastner1}  and Annalen der Physik~\cite{kastner2}.   After an overview of FET's and non-volatile FLASH memories, all exploiting the phenomenon of quantum mechanical tunneling, The SET came out as a neat consolidation of the ideas we discussed in class. The dot, schematically shown in Fig.~\ref{myth33}, has quantized energy levels but adding an electron costs charging energy $e^2/2C$, the so-called Coulomb blockade with $C$ being the capacitance of the dot. The dot is sandwiched between two metallic electrodes and separated by thin insulating layers. Electrons can tunnel into and out of the dot if they can `pay the price' of increased Coulomb energy which is achieved by pulling down all the levels of the dot by the application of a nearby positive potential. Furthermore spurious tunneling should also be curtailed and that happens if the uncertainty in charging energy is smaller than the energy itself. It ensures that electrons can tunnel onto and out of the dots only when they are asked to! Such a coherent description of the SET unified concepts in quantum tunneling, electrostatic energy and quantized energy levels in confined structures, besides showing the students how a real device exploited quantum physics.

\subsection{Imaging wavefunctions}

The king of at it all was undoubtedly the scanning tunneling microscope (STM) thoroughly discussed in textbooks. It was possible to showcase and explain an exhibition of unusually beautiful renderings of electron densities revealed in surface topographical images acquired from STM's~\cite{stm-gallery}.  Particularly interesting were the standing wave ripples produced in regions where the electrons were spatially confined or where they encountered potential energy discontinuities, either engineered or arising from defects introduced during device synthesis. I don't know how it transpired this way but as I was teaching this course, an incredible video also came hot off the press. The video aptly dubbed ``A Boy and his Atom" was constructed by researchers at IBM and comprised $220$ frames of STM images interlaced in quick succession~\cite{boy-atom}. The actors in this ``world's smallest movie" were oxygen atoms in carbon monoxide molecules skillfully dragged on a surface using surface forces in an STM. The movie was played in class and was an all-out hit!

\subsection{Quantum sniffing}

\begin{figure}[!t]
\begin{center}
\includegraphics[scale=0.4]{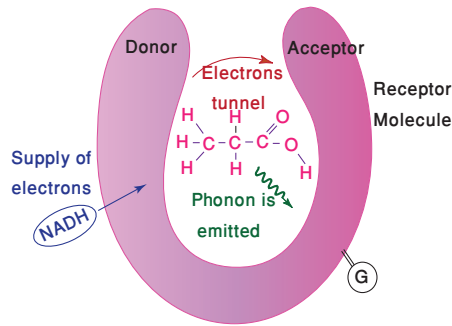}
\caption{Physical model of quantum sniffing of a fragrant molecule (ethyl acetate) explicable by phonon assisted tunneling. The NADH is a reducing agent and furnishes a supply of electrons. As electrons tunnel into the receptor region, the G-proteins are activated.
\label{myth3}}
\end{center}
\end{figure}

The STM culminated into describing, purportedly, how humans sniff and differentiate between different scents. The exact mechanism, though not yet fully understood, is believed to be based on inelastic tunneling of electrons~\cite{brookes} across a scent molecule which triggers the activation of cell-signaling G-proteins. Contrary to the elastic tunneling of electrons in an SET, in this case electrons lose energy as they tunnel across the scent molecule whose vibrational manifold has energy levels matching the exact decrease in electron energy~\cite{franco}. The tunneling is therefore enabled by the emission of phonons. Now the concept of phonons required a digression but this was rather easy as I had already introduced harmonic oscillators and quantized vibrational energy levels. In fact, I also took this as an opportunity to not only introduce the field of quantum biology~\cite{arndt}.

\section{Physics is a process, not a collection of results: real experiments shown in class\label{sec:demonstrations}}

A mainstay of the course were the nearly dozen demonstrations that were studded throughout the span of the course. These were brief experiments performed inside the classroom and capitulated during the formal lecture time. The underlying philosophy of including these demonstrations was quite simple.

The first goal was to make students remember physics. The Confucian adage from 250 BC, ``\emph{Tell me and I will forget; show me and I may remember; involve me, I will understand}" adds new dimensions to the already complicated teaching and learning exercise but highlights the role of visually appealing practical demonstrations and here I am \emph{not} talking about computer simulations. The idea behind the modern physics course for beginners, where an overwhelming majority of students were not intending physics majors, was to help them develop a scientifically sound and accurate appreciation of the inner workings of nature and not to heap upon them the heavy burden of insignificant technical intricacies---a burden they'll shed off with respite as they finally walk out of the examination hall on the finals day!

Second, I wanted to emphasize that physics (and science) is not an encyclopedia of facts, rather it's a process of discovery fueled by experiments which are the final arbiter attesting the `truths' of quantum mechanics. Ideas and tools~\cite{dyson} are the two tributaries feeding into the ocean of physics. Practical experiments, therefore, offer a glimpse into the process of scientific inquiry and discovery and help make quantum physics tangible and go a long way in showing that nothing of what we learn is a  figment in the imagination of some armchair thinkers or drawing room philosophers.

The third goal is exploring the connection with the history of science. Practical demonstrations, especially modern versions of historic experiments show that the development of quantum ideas doesn't happen in fits and starts, rather it's an evolutionary accumulation and we being students of physics in a city in Pakistan---the developing world---don't have to always feel like a hapless island but are part of a civilizational mainland and temporal continuum~\cite{donne}. We may be dwarfs but can definitely stand on the shoulders of giants (and can hence look further ahead)~\cite{bernard}.

What follows is a recounting of the several experiments I performed in class.

J.J. Thomson's classic experiment from 1897 on cathode rays~\cite{jjthomson} determining the $e/m$ ratio of electrons was reproduced using a scavenged cathode ray oscilloscope acquired from dump warehouses. The scope was opened up and the electron tube exposed after extending the electrical connections. Magnets could deflect the electron beam and could be applied in opposition to electric fields produced by vertical deflection plates.

The Nobel-Prize winning experiment of Franck and Hertz from $1914$ showed the existence of quantized energy levels in mercury atoms~\cite{franck-hertz}. The quantization was shown by the loss in energy of thermionically emitted electrons after discrete intervals of accelerating potential. These intervals corresponded to the first excitation potential of mercury. Using modern apparatus (Telatomic) that was interfaced with a computer running Labview enabled large-scale magnification on the projector screen. It was enjoyably possible to reproduce the original historic results and convince students of quantized energy manifolds.

\begin{figure}[!h]
\begin{center}
\includegraphics[scale=0.4]{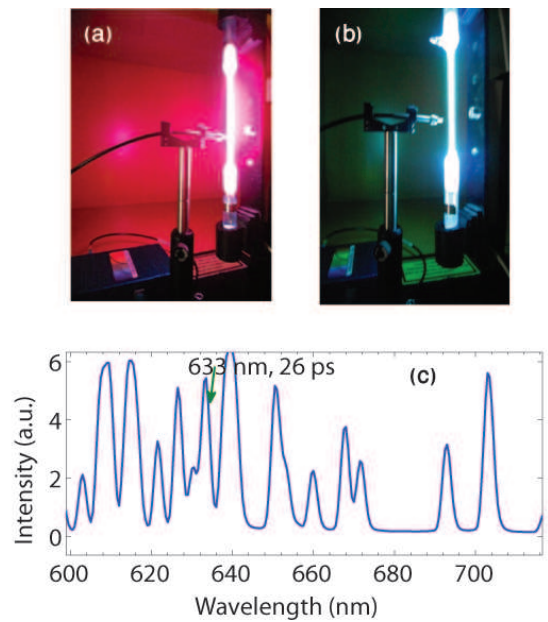}
\caption{(a) Pink glow from hydrogen, (b) bluish discharge from mercury and (c) line spectrum of neon discharge.
\label{demo1}}
\end{center}
\end{figure}

In my opinion another effective evidence of energy quantization was provided by directly looking at the line spectrum emitted from gaseous discharge. I lay my hands on a wonderful rugged miniature fiber optic spectrometer (Stellarnet) that quickly produced the spectrum of incoming light. With its help I was able to demonstrate spectrums from light bulbs, LED's and lasers. I then placed optical filters picking out narrow wavelength regimes and went on to show the spectrums from the colored discharge of hydrogen, neon and mercury (spectral emission tubes were from Pasco).

Towards the end of the course I could also measure linewidths from a mercury tube, demonstrating that the $633$ nm line was narrower and hence possessed a longer lifetime corresponding to a metastable state. This proved to be a much needed assurance for students who were always looking for some evidence of the energy-time uncertainty principle. The demonstrations with light sources of various kinds helped my students, in overall, realize that energy is quantized, light is composed of fields of various wavelengths that can be filtered out and finally, the peaks in a line spectrum have varying linewidths and intensities. Some discussion on the line intensities and their partial origin in Boltzmann distribution factors was also in place.

In order to show the property of coherence of light (as it was extremely important in my long drawn out discussions on interferometry) we built a Michelson interferometer and projected the characteristic circular pattern of fringes. Interferometry experiments were performed with a helium-neon laser (Thorlabs). I also showed diffraction patterns from single and double slits. The slit was made by breaking apart a razor blade and placing the broken halves back to back while for the double slit a common pin was inserted inside the gap.

\begin{figure}[!here]
\begin{center}
\includegraphics[scale=0.33]{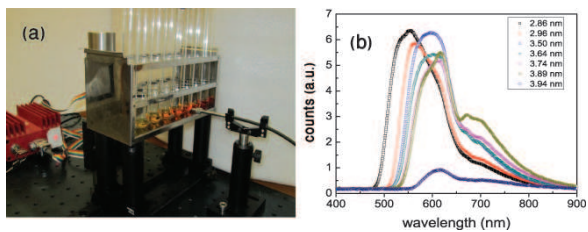}
\caption{(a) The arrangement for detecting the fluorescence of semiconductor quantum dots. The light source is a blue solid-state laser shown in the background and (b) fluorescence spectrums from quantum dots of various sizes showing the size-dependent quantization.
\label{demo3}}
\end{center}
\end{figure}

I also looked for some nice practical demonstration showing the effect of confinement on quantization. Students had derived the quantized energies of electrons of mass $m$ in an infinite well $E_n=n^2\pi^2\hbar^2/(2mL^2)$ of length $L$ and we prepared~\cite{nordell}, outside the class, cadmium selenide quantum dots of diameters varying in the range $2$ to $10$ nm. We used these quantum dots to illustrate the effect of varying the diameter on the separation between energy levels. Smaller sized dots lead to bigger gaps and hence the wavelengths are blue shifted. The different visual coloration of  these dots combined with quantitative spectral measurements were effective tools in demonstrating fluorescence and the role of ``size" in determining energy level spacings.

\begin{figure}[!here]
\begin{center}
\includegraphics[scale=0.4]{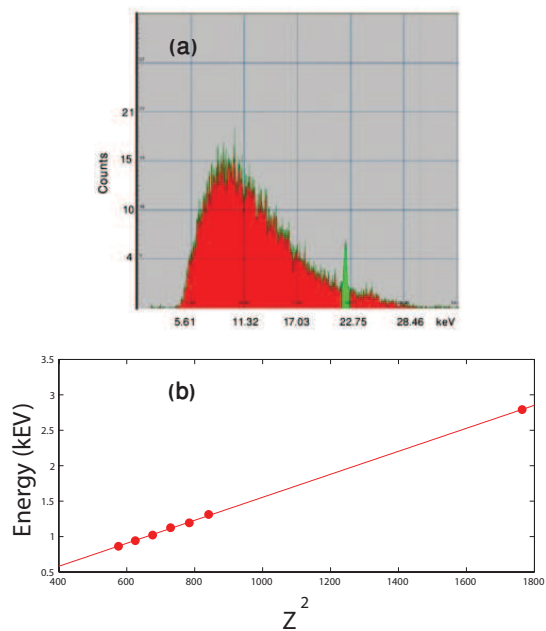}
\caption{(a) An X-ray spectrum showing bremmstrahlung radiation (in red) and characteristic X-rays (in green) (b) a verification of Moseley's law.
\label{demo4}}
\end{center}
\end{figure}

No introductory course on modern physics is complete without Bohr's model and other facets of the old quantum theory. Very early on in the course, students were convinced that energy is indeed quantized, but the last moment of truth dawned upon all of us when we used the characteristic X-rays emitted from the chromium in a steel sample to determine the ground state energy of the `hydrogenlike' atom. Using a home assembled X-ray fluorescence unit (individual components from Amptek) we analyzed the energy of chromium's $K_\alpha$ peak and after accounting electron-shielding resulting in an effective nuclear charge of chromium's atomic number less one, we arrived at an energy estimate really close to $13.6~$eV. The fluorescence unit also showed really nice visual proofs of Bremmstrahlung as well as characteristic X-rays from different elements whose energies depended on atomic numbers (in concordance with Moseley's predictions~\cite{moseley}). Stainless steel with an assortment of iron, nickel, chromium and cobalt was quite a suitable choice for these fluorescence experiments.

The Prutchi family recently wrote a book on demonstrating quantum physics through hands-on projects~\cite{prutchi}. They demonstrated tunneling using microwave radiation but the effect cannot be seen. I chose instead to use the visually appealing effect of the tunneling optical radiation from a bright green laser pointer. The light fell on the hypotenuse of a right angled prism at an angle greater than the critical angle. Obviously the light was totally internally reflected but bringing another prism close by, the total internal reflection could be frustrated, and the evanescent wave could be picked up and transmitted~\cite{you}.

In order to make students appreciate the statistical nature of tunneling, and in retrospect, all, quantum processes, I also set up an experiment employing a radioactive source and a Geiger Muller tube interfaced with the computer picking up decay events. The experiment was started at the beginning of one lecture and was left running on its own. Towards the end of the class a beautiful Poisson distribution of radioactive decays had built up.

In the end, of course, I had to show electron diffraction and this came from a prefabricated electron diffraction tube (3BScientific) in which a fine beam of electrons is diffracted from a polycrystal of graphite deposited on nickel. The diffraction rings were the last items shown in the class verifying that electrons are indeed a field.

\subsection{Demonstrations: how not do them}

Based on student feedback, I realized that these practical experiments were an important tool in enhancing student understanding  of the subject, but there are also words of caution. First, demonstrations are like life-saving drugs that can also turn fatal if not administered properly. One needs to practice these experiments and be engrossed in their development offline. There is not a bigger disaster than walking into the classroom with a borrowed demonstration kit and ending up in a wild goose search for the power-up button or hastily rummaging through the product's user manuals eating into premium lecturing time. If you want to do a similar experiment for your classes, my advice is to follow it up with full vigor and dedication and, of course do lots of practice.

Second, a practical experiment performed in class can easily become a Trojan horse if it doesn't disseminate to the entire class. What good is a demonstration on tunneling if the back-benchers can't see the beam of light, or if the cameraman is obscuring the view of fringes in an interferometry experiment or when the spectral lines haven't been projected with sufficient magnification to make the fringes clearly visible! I strongly recommend that these demonstrations should be carefully sited inside the lecture hall. After all it's always good to make a sketch of the layout before hand.

The timing and duration of these demonstrations is also crucial for these should not pass away like fleeting angels nor be so boring and dragging that students lose attention. Best are the experiments whose outcomes are clear, visually observable and require hardly any or little post-processing.

Another pitfall I tried to protect against was not to pose these demonstrations as mere entertaining jugglery. These were serious activities upon which students were assessed in exams, the equipment was elegantly and aesthetically designed and all safety precautions, for example, wearing safety goggles while operating bright lasers, insulation gloves in the operation of a high voltage power supply for demonstrating field emission and reminding to students the importance of radiation safety while demonstrating radioactivity, gave a solemn outlook to these demonstrations. I also found that despite these sobering precautions the jaw-dropping excitement factor remains very much intact.

\section{Student feedback and assessment\label{sec:feedback}}

\begin{figure}[!here]
\begin{center}
\includegraphics[scale=0.5]{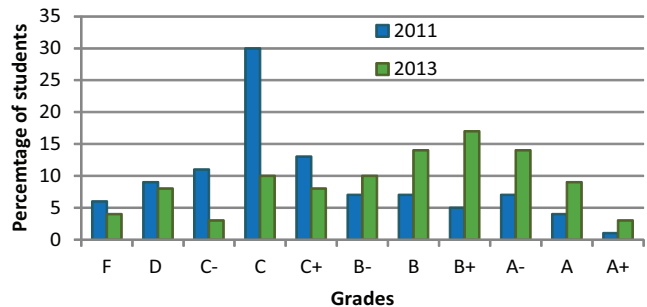}
\caption{Comparison between student performance in 2011 and 2013 showing a histogram of student grades.
\label{comparison}}
\end{center}
\end{figure}

Figure~\ref{comparison} shows students' consolidated grades after the conclusion of the course offered in 2013 alongside the grades in 2011 when I offered a similar course from a standard text~\cite{serway}, taught rather conventionally, and which did not include any in-class demonstrations. The improvement in class averages is quite evident. The student class sizes were similar and the grade thresholds are identical to within $\pm 2\%$ of the percentage marks. The modal grade went up from C to B+. Students seemed to be excited about physics, and claimed to have learnt lots of new physics at the conclusion of the course. Out of 110 students who filled in the post-course survey, responding to the statement, \emph{``I have learned a lot of new things by taking this course,''}, 95 responded with ``Excellent'', 12 with ``Good'', 2 with ``Good'', 1 with ``Satisfactory'' and 0 with ``Unsatisfactory''. Selected anonymous student feedback is reproduced below.

\begin{quote}
\textbf{S1} ``The instructor managed to make a subject like Physics a lot of fun and regularly made sure that we knew things in the larger context of the world; how technology fills in the jigsaw puzzle.''
\end{quote}

\begin{quote}
\textbf{S2} ``Presentations of experiments and videos of simulations made the lecture all the more interesting''
\end{quote}

\begin{quote}
\textbf{S3}``This, modern physics, course was the best course that I took in my freshman year at SSE. The experiments that were shown in the class greatly helped in developing my concepts.''
\end{quote}

\begin{quote}
\textbf{S4}``Without a doubt the experiments that were performed in class [were some of the strengths of this course].''
\end{quote}

\section{Conclusions\label{sec:conclusions}}

In conclusion, I would like to share some overall lessons I have learned through the teaching of quantum physics with the hope that some of these tips could be useful for other teachers of similar level courses. Very briefly, with the poor educational background of our students entering university, yet their harboring of fanciful notions of being smart~\cite{courtney}, it is sometimes important to make them unlearn and then learn. Concepts stuck from high-school sometimes come in the way of the kind of learning I have emphasized here. Sticking to particles as the real thing will hamper the assimilation of the concept of quantized fields, insisting that entropy is disorder hides the notion of energy dispersal and so on. Furthermore, formula-orientedness of our students who consider that formulas drive concepts and ideas is a dangerous attitude and must be challenged right from the beginning.

Second, my experience has been employing lots and lots of sketches on the blackboard. This habit makes physics less sketchy. It also invites conceptual clarity and rigor for the teacher as well as the students. It is also important to have students make sketches and plot functions right inside the classroom for there is no better way of learning what a mathematical equation means than \emph{plotting} its solution. Plot fields $\Psi(x,t)$ frozen in time and frozen in space; plot real and imaginary parts of fields, and also plot their absolute values. Make students also plot the fields for potentials of different kinds such as triangular potentials, semi-infinite wells but also make them plot the potentials for a variety of physical scenarios. Bouncing quantum balls~\cite{banacloche} and falling atoms are beautiful physical experiments to test these capabilities of students.  And yes, don't forget to emphasize the time-dependent part of fields preferably through interactive computer generated plots in the fashion of the reference~\cite{singh}.

\section{Appendix: Selected questions}
\begin{enumerate}
\item Figures (a) through (f) show various kinds of potential steps and obstacles to an electron injected  from the left, with energy $E$. $V(x)$ is shown by solid lines and $E$ by dashed lines. Two or three regions (I, II and III) are also identified. In each case, discuss the following.
    (i) Fields (wavefunctions) in each region---their mathematical form and sketches of their real parts.
    (ii) Identify the discontinuities from where reflection of the single electron can take place.
    \begin{center}
    \includegraphics[scale=0.5]{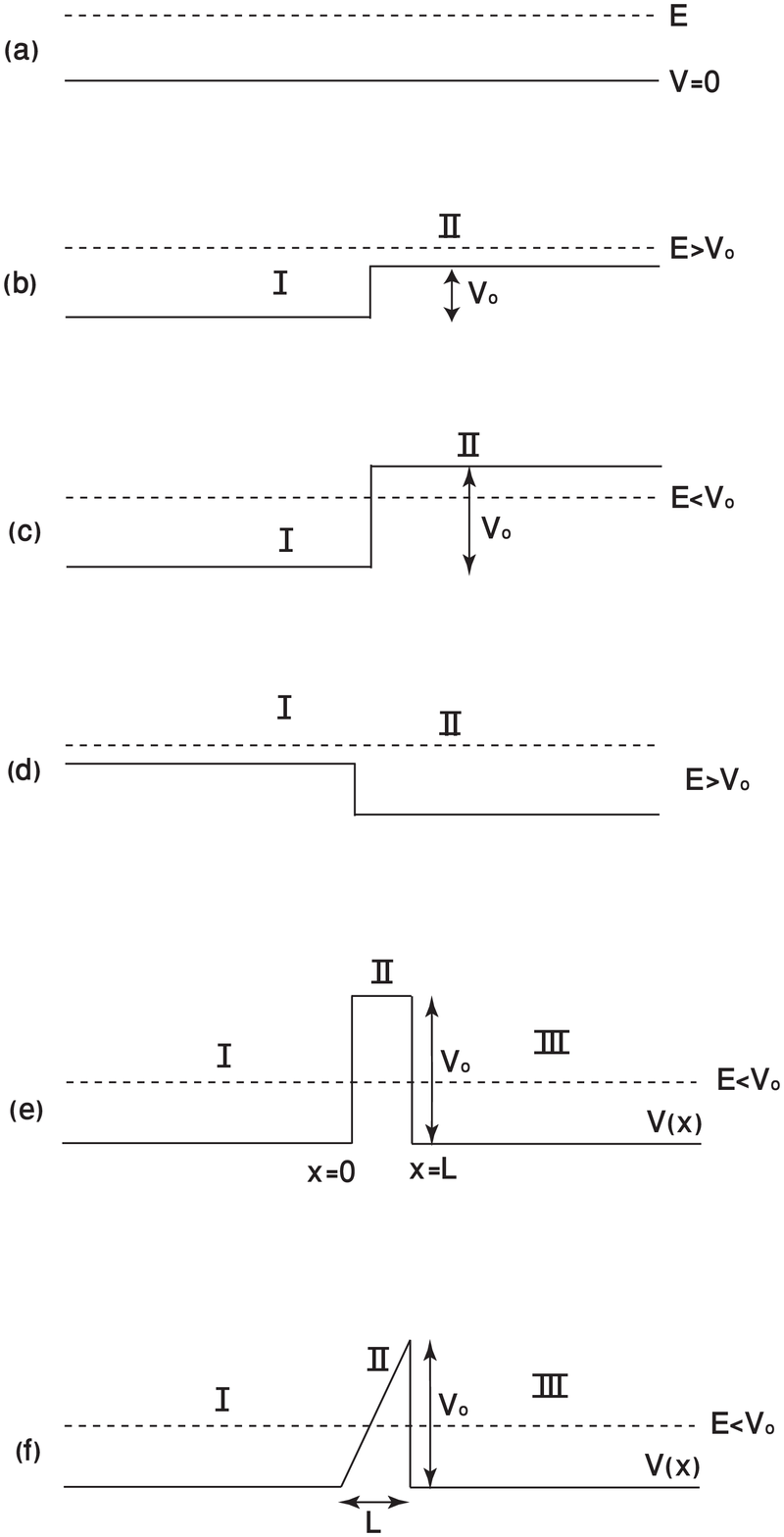}
    \end{center}

\item There are probably $10^{40}$ oxygen molecules in this auditorium. They are evenly spaced. If all of these molecules were to spontaneously congregate in one small corner, all of us would probably suffocate. When is this spontaneous congregation of molecules more likely?

    \begin{center}
    \includegraphics[scale=0.5]{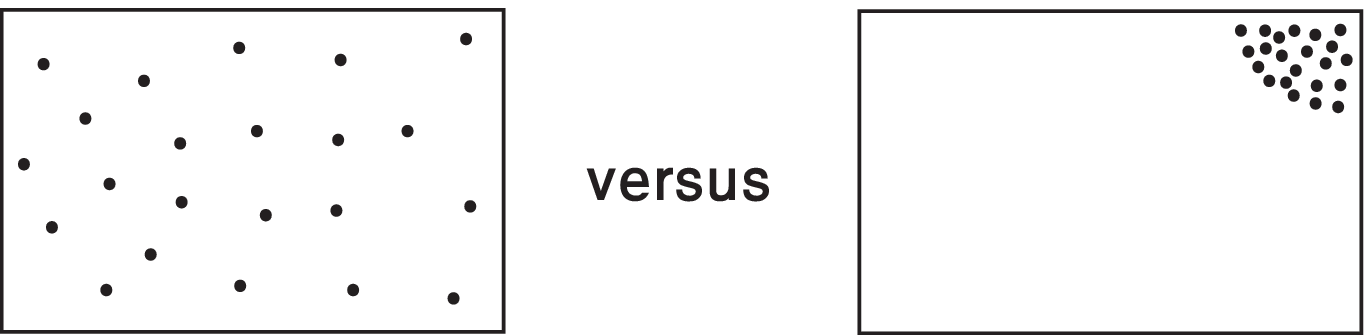}
    \end{center}

    (a) When there is a higher number of $O_2$ molecules.\\
    (b) When there are fewer $O_2$ molecules.\\
    (c) The likelihood of this congregation of molecules is independent of the number of molecules.\\
    (d) By increasing the pressure of $O_2$ inside the auditorium.\\
    (e) None of the above.\\

\item Suppose you are provided with many identically prepared electrons. You perform multiple experiments, one electron at a time. The goal is to find the ``electron's position". Your multiple experiments reveal a distribution as shown below. The distribution is centered around $x_0$.
    \begin{center}
    \includegraphics[scale=0.3]{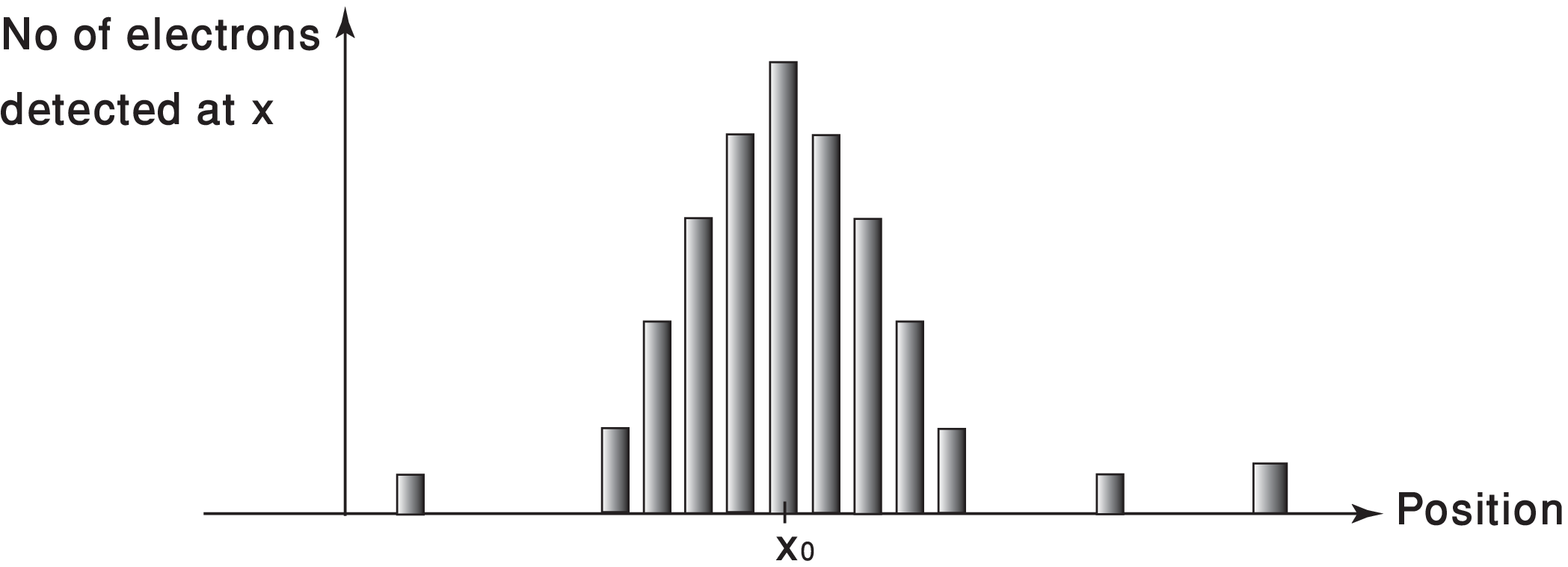}
    \end{center}
    What can be said, {\bf{most accurately}}, about the momentum of electron?\\
    (a) Each electron has a precise but different momentum.\\
    (b) Each electron has a different momentum.\\
    (c) It is impossible to define the momentum of an electron.\\
    (d) Each electron has a similar range of possibilities of momentum.\\
    (e) The measuring instrument measures the location, hence we cannot say absolutely anything about the momentum.
\end{enumerate}


\begin{thebibliography}{99}
\bibitem{bernard}{The quotation is attributed to the French philosopher Bernard of Chartres from the thirteenth century.}
\end{thebibliography}
\end{document}